%% file: ke2_.tex
\documentclass[11pt]{article}
\usepackage[dvips]{graphicx}
\usepackage{epsfig}
\usepackage{amssymb}
\usepackage{color}
\usepackage[left]{lineno}

\begin{document}
\centerline{\LARGE EUROPEAN ORGANIZATION FOR NUCLEAR RESEARCH}
%
%
\vspace{20mm} {\flushright{
CERN-PH-EP-2011-004 \\
21 January 2011
\\}}
\vspace{-35mm}
%
%
\vspace{35mm}

\begin{center}
{\bf {\Large \boldmath{Test of Lepton Flavour Universality in
$K^+\to\ell^+\nu$ Decays}}}
\end{center}
\begin{center}
{\Large The NA62 collaboration$\,$\renewcommand{\thefootnote}{\fnsymbol{footnote}}%
\footnotemark[1]\renewcommand{\thefootnote}{\arabic{footnote}}}\\
\end{center}

\begin{abstract}
A precision test of lepton flavour universality has been performed
by measuring the ratio $R_K$ of kaon leptonic decay rates $K^+\to
e^+\nu$ and $K^+\to\mu^+\nu$ in a sample of 59813 reconstructed
$K^+\to e^+\nu$ candidates with $(8.71\pm0.24)\%$ background
contamination. The result $R_K=(2.487\pm0.013)\times10^{-5}$ is in
agreement with the Standard Model expectation.
\end{abstract}

\begin{center}
\it{Accepted for publication in Physics Letters B}
\end{center}

\setcounter{footnote}{0}
\renewcommand{\thefootnote}{\fnsymbol{footnote}}
\footnotetext[1]{Copyright CERN for the benefit of the NA62
collaboration. Contact: Evgueni.Goudzovski@cern.ch.}
\renewcommand{\thefootnote}{\arabic{footnote}}

\newpage
\include{na62-authors-2007}
\newpage


\section*{Introduction}

In the Standard Model (SM) the decays of pseudoscalar mesons to
light leptons are helicity suppressed.
In particular, the SM width of $P^\pm\to\ell^\pm\nu$
decays with $P=\pi,K,D,B$ (denoted $P_{\ell 2}$ in the following) is
\begin{equation}
\Gamma^\mathrm{SM}(P^\pm\to\ell^\pm\nu) = \frac{G_F^2 M_P
M_\ell^2}{8\pi} \left(1-\frac{M_\ell^2}{M_P^2}\right)^2
f_P^2|V_{qq\prime}|^2, \label{eq:sm}
\end{equation}
where $G_F$ is the Fermi constant, $M_P$ and $M_\ell$ are meson and
lepton masses, $f_P$ is the decay constant, and $V_{qq\prime}$ is
the corresponding Cabibbo-Kobayashi-Maskawa matrix element. Although
the SM predictions for the $P_{\ell 2}$ decay rates are limited by
hadronic uncertainties, their specific ratios do not depend on $f_P$
and can be computed very precisely. In particular, the SM prediction
for the ratio $R_K=\Gamma(K_{e2})/\Gamma(K_{\mu 2})$ of kaon
leptonic decay widths inclusive of internal bremsstrahlung (IB)
radiation is~\cite{ci07}
\begin{equation}
\label{Rdef} R_K^\mathrm{SM} = \left(\frac{M_e}{M_\mu}\right)^2
\left(\frac{M_K^2-M_e^2}{M_K^2-M_\mu^2}\right)^2 (1 + \delta
R_{\mathrm{QED}})= (2.477 \pm 0.001)\times 10^{-5},
\end{equation}
where $\delta R_{\mathrm{QED}}=(-3.79\pm0.04)\%$ is an
electromagnetic correction due to the IB and structure-dependent
effects.

Within certain two Higgs doublet models (2HDM),
including the minimal supersymmetric model (MSSM), $R_K$ is
sensitive to lepton flavour violating (LFV) effects appearing at the
one-loop level via the charged Higgs boson ($H^\pm$)
exchange~\cite{ma06,ma08}, representing a unique probe into mixing
in the right-handed slepton sector~\cite{el09}. The dominant
contribution due to the LFV coupling of the $H^\pm$ is
\begin{equation}
R_K^\mathrm{LFV}\simeq
R_K^\mathrm{SM}\left[1+\left(\frac{M_K}{M_H}\right)^4
\left(\frac{M_\tau}{M_e}\right)^2 |\Delta
_R^{31}|^2\tan^6\beta\right], \label{rk_lfv}
\end{equation}
where $\tan\beta$ is the ratio of the two Higgs vacuum expectation
values, and $|\Delta_{R}^{31}|$ is the mixing parameter between the
superpartners of the right-handed leptons, which can reach $\sim
10^{-3}$. This can enhance $R_K$ by ${\cal O}(1\%)$ without
contradicting any experimental constraints known at present,
including upper bounds on the LFV decays $\tau\to eX$ with
$X=\eta,\gamma,\mu\bar\mu$. On the other hand, $R_K$ is sensitive to
the neutrino mixing parameters within the SM extension involving a
fourth generation~\cite{la10}.

The first measurements of $R_K$ were performed in the
1970s~\cite{cl72, he75, he76}; the current PDG world
average~\cite{pdg} is based on a more precise recent
result~\cite{am09} $R_K=(2.493\pm0.031)\times 10^{-5}$. A new
measurement of $R_K$ based on a part of the data sample collected by
the NA62 experiment at CERN in 2007 is reported in this letter. The
analyzed $K_{e2}$ sample is $\sim 4$ times larger than the total
world sample, allowing a measurement of $R_K$ with a precision well
below 1\%.

\section{Beam, detector and data taking}

The beam line and setup of the NA48/2 experiment~\cite{fa07, ba07}
have been used for the NA62 data taking in 2007. Experimental
conditions have been optimized for the $K_{e2}/K_{\mu 2}$
measurement.

\subsection{Kaon beam}

The beam line was originally designed to deliver simultaneous
unseparated $K^+$ and $K^-$ beams derived from the primary 400
GeV/$c$ protons extracted from the CERN SPS. In 2007, the muon
sweeping system was optimized for the positive beam, and the sample
used for the present analysis was collected with the $K^+$ beam
only. Positively charged particles within a narrow momentum band
with a central momentum of 74.0~GeV/$c$ and a spread of 1.4~GeV/$c$
(rms) are selected by the first two magnets in a four dipole
achromat and by momentum-defining slits incorporated into a 3.2~m
thick copper/iron proton beam dump, which also blocks the
negatively-charged particles. The beam subsequently passes through
acceptance-defining and cleaning collimators and a set of four
quadrupoles of alternating polarity, as well as muon sweeping
magnets, before entering the fiducial decay volume contained in a
114~m long cylindrical vacuum tank with a diameter of 1.92~m
upstream, increasing to 2.4~m downstream.

With about $1.8\times 10^{12}$ primary protons incident on the
target per SPS pulse of about 4.8~s duration repeating every 14.4 or
16.8~s, the secondary beam flux at the entrance of the decay volume
is $2.5\times 10^7$ particles per pulse. The fractions of $K^+$,
$\pi^+$, $p^+$, $e^+$ and $\mu^+$ in the secondary beam are 0.05,
0.63, 0.21, 0.10 and 0.01, respectively. The fraction of those beam
kaons decaying in the vacuum tank at nominal momentum is $18\%$. The
beam transverse size at the entrance to the decay volume is $\delta
x = \delta y = 4$~mm (rms), and its horizontal and vertical angular
divergences are about $20~\mu{\rm rad}$ (rms). The beam central
momentum, transverse position at the entrance to the vacuum tank and
direction varied slowly over time with respect to the nominal ones
in the ranges of $\sim 0.1~{\rm GeV}/c^2$, $\sim 1~{\rm mm}$ and
$\sim 10~\mu{\rm rad}$, respectively.

The beam line also transmits certain off-momentum charged kaons and
pions punching through the proton beam dump with a suppression
factor of $\sim 10^{-3}$. However the subsequent decays of these
particles do not contribute to the signal region of the present
analysis.

\subsection{Detector}

The charged particle properties are measured in a magnetic
spectrometer, housed in a tank filled with helium at nearly
atmospheric pressure, placed after the decay volume and separated
from the vacuum by a thin ($\sim 0.4\%$ radiation lengths $X_0$)
$\rm{Kevlar}\textsuperscript{\textregistered}$ window. The
spectrometer comprises four drift chambers (DCHs)~\cite{be95}, two
upstream and two downstream of a dipole magnet which gives a
horizontal transverse momentum kick of $265~\mathrm{MeV}/c$ to
singly-charged particles. Each DCH is composed of eight planes of
sense wires, and provides a spatial resolution of $90~\mu$m in each
projection. The measured momentum resolution is $\sigma_p/p = 0.48\%
\oplus 0.009\%\cdot p$, where $p$ is expressed in GeV/$c$.

A plastic scintillator hodoscope (HOD) producing fast trigger
signals and providing precise time measurements of charged particles
is placed after the spectrometer. It consists of a plane of vertical
strips, followed by a similar plane of horizontal strips (128
counters in total). Both planes have regular octagonal shapes and a
central hole for the passage of the beam.

The HOD is followed by a quasi-homogeneous liquid krypton
electromagnetic calorimeter (LKr)~\cite{ba96} used for lepton
identification and as a photon veto in the present analysis. The LKr
is 127~cm (or $27X_0$) thick along the beam, with projective readout
consisting of copper/beryllium ribbons extending from the front to
the back of the detector. The 13248 readout cells have a transverse
size of approximately 2$\times$2 cm$^2$ each and have no
longitudinal segmentation. The energy resolution is
$\sigma_E/E=0.032/\sqrt{E}\oplus0.09/E\oplus0.0042$ ($E$ in GeV).
The spatial resolution for the transverse coordinates $x$ and $y$ of
an isolated electromagnetic shower is
$\sigma_x=\sigma_y=0.42/\sqrt{E}\oplus0.06$ cm ($E$ in GeV).

An aluminium beam pipe of 158~mm outer diameter and 1.1~mm thickness
traversing the centres of all detector elements allows the undecayed
beam particles to continue their path in vacuum. The outer
transverse sizes of the subdetectors are about 2.4~m.

\newpage
\subsection{Trigger and data acquisition}

A minimum bias trigger configuration has been employed, resulting in
high efficiency. The $K_{e2}$ trigger condition consists of
coincidences of signals in the two HOD planes (the $Q_1$ signal),
loose lower and upper limits on DCH hit multiplicity (the 1-track
signal), and LKr energy deposit $(E_\mathrm{LKr})$ of at least 10
GeV. The $K_{\mu 2}$ trigger condition requires a coincidence of the
$Q_1$ and 1-track signals downscaled by a factor $D=150$.
The non-downscaled $K_{\mu 2}$ trigger rate is
0.5~MHz, and is dominated by beam halo muons; the $K_{e2}$ trigger
rate is about 10~kHz. Downscaled control samples based on trigger
signals from the DCHs, HOD and LKr have been collected to monitor
the performance of the main trigger signals. The data taking took
place during four months starting in June 2007. About 40\% of the
350k recorded good SPS spills are used for the present analysis.

\section{Analysis strategy}

The analysis strategy is based on counting the numbers of
reconstructed $K_{e2}$ and $K_{\mu 2}$ candidates collected
concurrently. Therefore the analysis does not rely on an absolute
beam flux measurement, and several systematic effects (due to beam
simulation, accidental activity, charged track reconstruction, $Q_1$
trigger efficiency, and time-dependent effects) cancel at first
order.

Due to the significant dependence of acceptance and background on
lepton momentum, the $R_K$ measurement is performed independently in
10 momentum bins covering a range from 13 to 65~GeV/$c$. The lowest
momentum bin spans 7 GeV/$c$, while the others are 5 GeV/$c$ wide.
The selection criteria have been optimized separately in each
momentum bin. The data samples in the momentum bins are
statistically independent, however the systematic errors are
partially correlated. The ratio $R_K$ in each bin is computed as
\begin{equation}
R_K = \frac{1}{D}\cdot \frac{N(K_{e2})-N_{\rm B}(K_{e2})}{N(K_{\mu
2}) - N_{\rm B}(K_{\mu 2})}\cdot \frac{A(K_{\mu 2})}{A(K_{e2})}
\cdot \frac{f_\mu\times\epsilon(K_{\mu 2})}
{f_e\times\epsilon(K_{e2})}\cdot\frac{1}{f_\mathrm{LKr}},
\label{RKexp}
\end{equation}
where $N(K_{\ell 2})$ are the numbers of selected $K_{\ell 2}$
candidates $(\ell=e,\mu)$, $N_{\rm B}(K_{\ell 2})$ are the numbers
of background events, $A(K_{\mu 2})/A(K_{e2})$ is the ratio of the
geometric acceptances (referred to as the acceptance correction in
the following), $f_\ell$ are the lepton identification efficiencies,
$\epsilon(K_{\ell 2})$ are the trigger efficiencies,
$f_\mathrm{LKr}$ is the global efficiency of the LKr readout (which
affects only the $K_{e2}$ selection), and $D=150$ is the $K_{\mu 2}$
trigger downscaling factor.

To evaluate the acceptance correction and the geometric parts of the
acceptances for background processes entering the computation of
$N_B(K_{\ell 2})$, a detailed Monte Carlo (MC) simulation based on
Geant3~\cite{geant3} is used. It includes a description, with time
variations, of the beam line optics, the full detector geometry,
materials, magnetic fields, local inefficiencies of DCH wires, and
inactive LKr cells (0.8\% of channels). Particle identification,
trigger and readout efficiencies are measured directly from data.

\section{Data analysis}
\subsection{Event reconstruction and selection}
\label{sec:selection}

Charged particle trajectories are reconstructed from hits and drift
times in the spectrometer. Track momenta are evaluated using a
detailed magnetic field map. Fine calibrations of spectrometer field
integral and DCH alignment are performed by monitoring the mean
reconstructed $K^+\to\pi^+\pi^+\pi^-$ invariant mass, and the
missing mass in $K_{\mu 2}$ decays.

Clusters of energy deposition in the LKr are found by locating the
maxima in the digitized pulses from individual cells in both space
and time and accumulating the energy within a radius of
approximately 11~cm. Shower energies are corrected for energy
outside the cluster boundary, energy lost in inactive cells and
cluster energy sharing. The energy response has been calibrated with
samples of positrons from $K^+\to\pi^0e^+\nu$ decays.

Due to the topological similarity of $K_{e2}$ and $K_{\mu 2}$
decays, a large part of the selection is common for the two decay
modes, which leads to significant cancellations of the related
systematic uncertainties. The main selection criteria are listed
below.
\begin{itemize}
\item Exactly one reconstructed charged particle track (lepton
candidate) geometrically consistent with originating from a kaon
decay is required. The electric charge of the track must be
positive.
\item The extrapolated track impact points in the DCHs, HOD and
LKr must be within their geometrical acceptances. The LKr acceptance
condition includes appropriate separations from the detector edges
and inactive cells.
\item The reconstructed track momentum must be in the range
13 to 65~GeV/$c$. The lower limit ensures high efficiency of the
$E_\mathrm{LKr}>10$~GeV trigger condition. Above the upper limit,
the analysis is affected by large uncertainties due to background
subtraction.
\item No LKr clusters with energy $E>E_\mathrm{veto}=2$~GeV and in time
with the track are allowed, unless they can be associated to the
track via direct energy deposition or bremsstrahlung. (Most clusters
due to bremsstrahlung in front of the spectrometer magnet are
resolved from those directly deposited by the track). This
requirement provides a photon veto for suppression of backgrounds
from $K^+\to e^+\nu\gamma$, $K^+\to\pi^0e^+\nu$, and
$K^+\to\pi^+\pi^0$ decays. However the veto is not hermetic due to
the beam pipe and the limited transverse size of the LKr.
\item The decay vertex is reconstructed as the point of closest approach
of the lepton candidate track extrapolated upstream, and the kaon
beam axis. The measured stray magnetic field in the vacuum tank is
taken into account. The position of the kaon beam axis is monitored
with a sample of fully reconstructed $K^+\to\pi^+\pi^+\pi^-$ decays.
\item The distance from the kaon decay vertex to the beginning of the
vacuum tank is required to exceed a minimum value ranging from 8~m
at low lepton momentum to 43~m at high momentum, which removes the
bulk of the beam halo background (discussed in
Section~\ref{sec:halo}).
\item For further suppression of the beam halo and several other backgrounds,
the reconstructed closest distance of approach of the track to the
beam axis must not exceed 3.5~cm.
\end{itemize}
The following two main criteria are used to distinguish $K_{e2}$
from $K_{\mu 2}$ decays.
\begin{itemize}
\item The kinematic identification of $K_{e2}$ ($K_{\mu 2}$) decays
is based on constraining the reconstructed squared missing mass in
the positron (muon) hypothesis:
\begin{equation}
-M_1^2 < M_{\mathrm{miss}}^2(\ell) = (P_K - P_\ell)^2 < M_2^2.
\end{equation}
Here $P_K$ is the average kaon four-momentum (monitored in time with
$K^+\to\pi^+\pi^+\pi^-$ decays), and $P_\ell$ is the reconstructed
lepton four-momentum (under the positron or muon mass hypothesis).
The limits $M_1^2$ and $M_2^2$ have been optimized for each lepton
momentum bin, taking into account the $M_{\mathrm{miss}}^2(\ell)$
resolution (which varies from 0.0025 (GeV/$c^2$)$^2$ at mid track
momentum to 0.005 (GeV/$c^2$)$^2$ at low and high track momentum),
the radiative mass tails, and the background conditions. $M_1^2$
varies between 0.013 and 0.016 (GeV/$c^2$)$^2$ and $M_2^2$ between
0.010 and 0.013 (GeV/$c^2$)$^2$. The kinematic separation of
$K_{e2}$ and $K_{\mu2}$ decays is illustrated in
Fig.~\ref{fig:ke2-km2-kinematic-separation}a.
\item The lepton identification is based on the ratio $E/p$ of
energy deposition in the LKr to momentum measured by the
spectrometer. Charged particles with $(E/p)_\mathrm{min}<E/p<1.1$,
where $(E/p)_\mathrm{min}=0.95$ for $p>25$~GeV/$c$ and
$(E/p)_\mathrm{min}=0.9$ otherwise, are identified as positrons. At
low lepton momenta, the background from particle mis-identification
is negligible. For $p>25$~GeV/$c$, the larger $(E/p)_\mathrm{min}$
limit minimises the net uncertainty from $K_{\mu 2}$ background
subtraction and particle mis-identification inefficiency. Charged
particles with $E/p<0.85$ are classified as muons. The data $E/p$
spectra of positrons and muons are shown in
Fig.~\ref{fig:ke2-km2-kinematic-separation}c.
\end{itemize}

\boldmath
\subsection{The $K_{e2}$ sample}
\unboldmath

The number of $K_{e2}$ candidates in the signal region is
$N(K_{e2})=59813$. The sources of background in the $K_{e2}$ sample
are discussed below.

\boldmath \subsubsection{$K_{\mu2}$ background} \unboldmath

\begin{figure}[tb]
\begin{center}
\resizebox{0.50\textwidth}{!}{\includegraphics{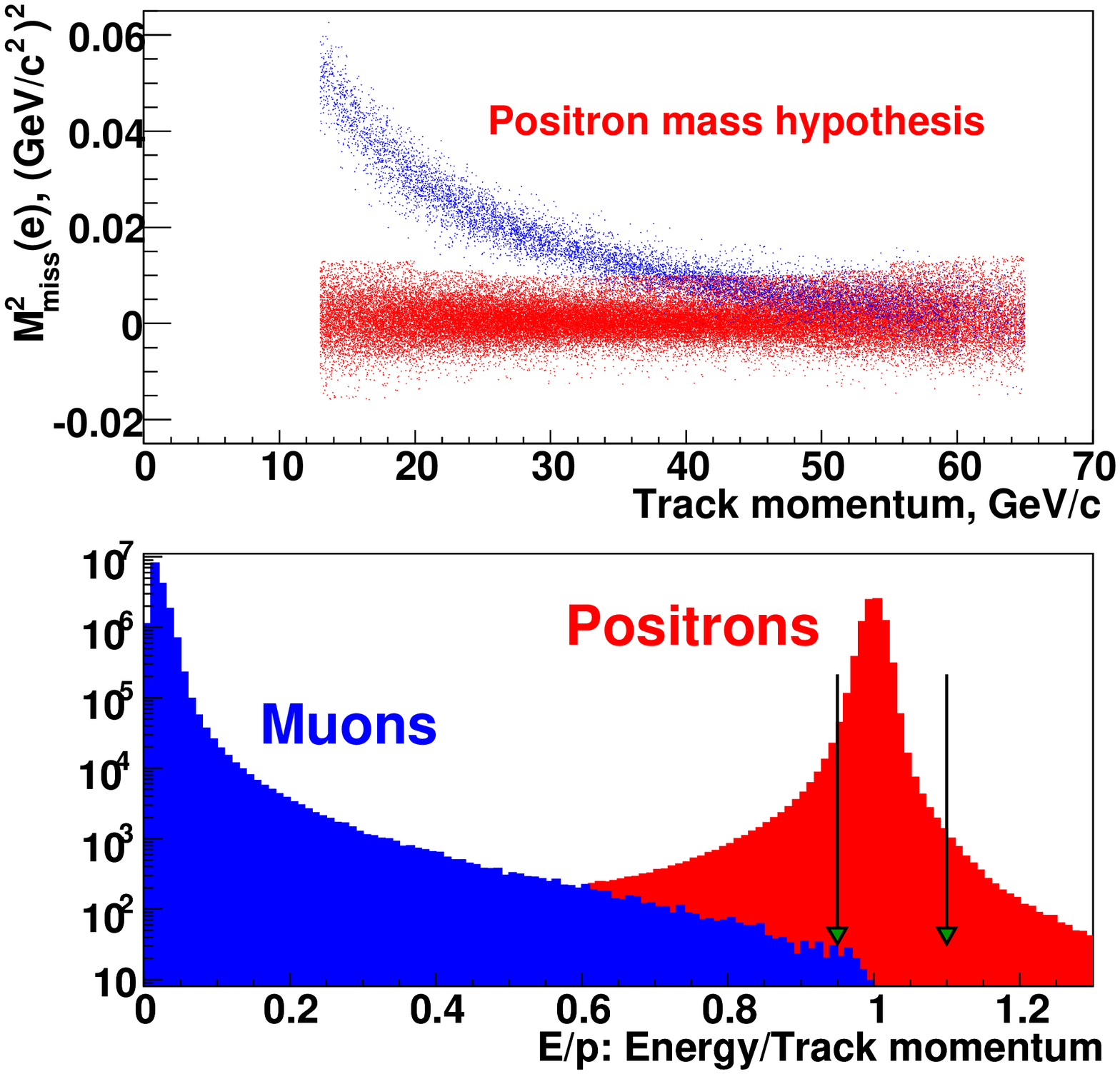}}%
\resizebox{0.50\textwidth}{!}{\includegraphics{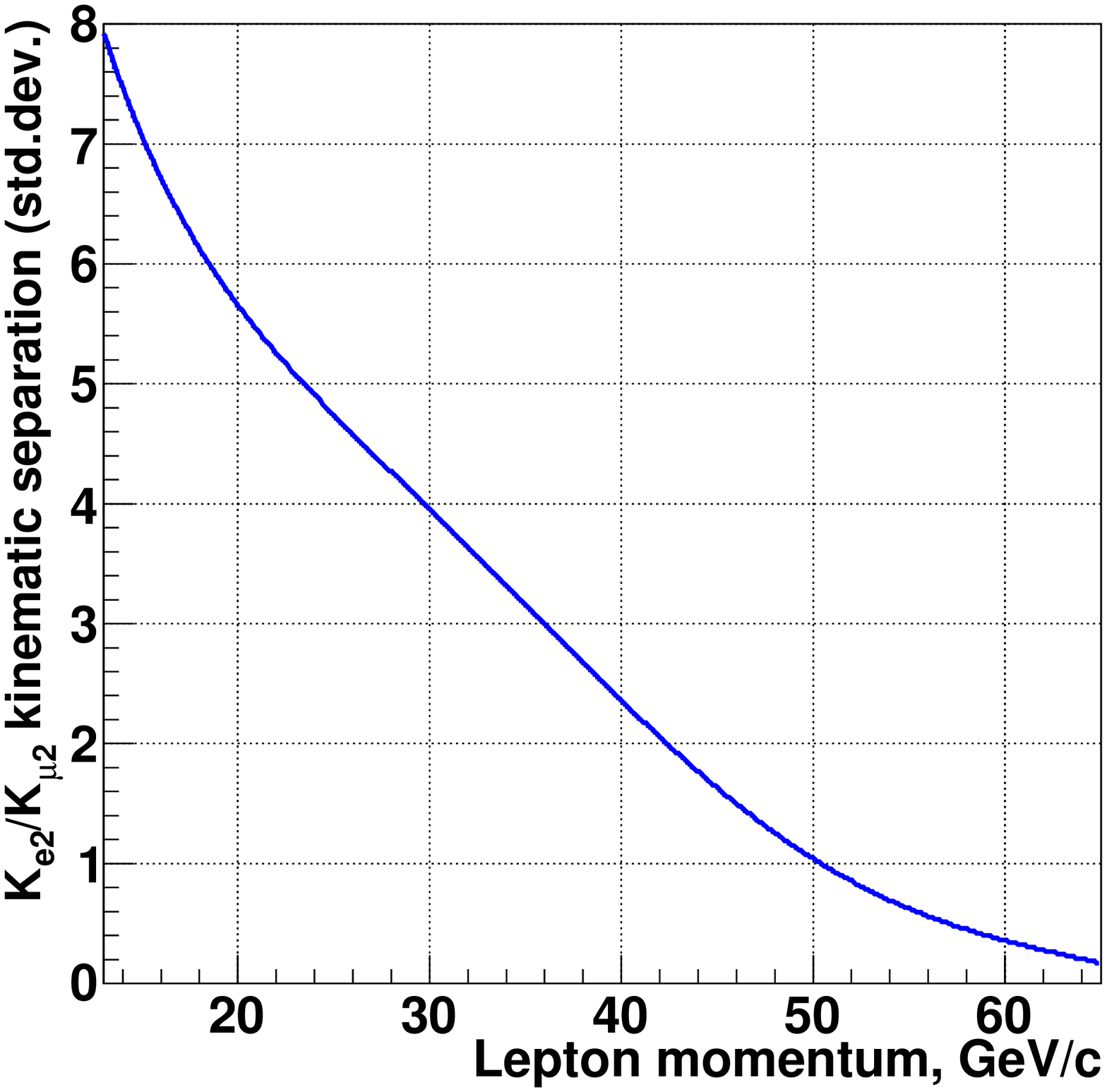}}
\put(-255,200){\bf\large (a)} \put(-255,91){\bf\large (c)}
\put(-25,198){\bf\large (b)} \put(-413,148){\color{red}$K_{e2}$}
\put(-413,195){\color{blue}$K_{\mu2}$}
\end{center}
\vspace{-12mm} \caption{(a) Squared missing mass assuming the
positron mass hypothesis $M^2_{\rm miss}(e)$ as a function of lepton
momentum for reconstructed $K_{e2}$ and $K_{\mu2}$ decays (data);
(b) $K_{e2}$ vs $K_{\mu 2}$ kinematic separation (standard
deviations) as a function of lepton momentum; (c) $E/p$ spectra of
positrons and muons (data); the positron identification limits for
$p>25$~GeV/$c$ are indicated by arrows.}
\label{fig:ke2-km2-kinematic-separation}
\end{figure}

Kinematic separation of $K_{e2}$ from $K_{\mu 2}$ decays is
achievable at low lepton momentum only ($p\lesssim 35$~GeV/$c$), as
shown in Figs.~\ref{fig:ke2-km2-kinematic-separation}a and
\ref{fig:ke2-km2-kinematic-separation}b. At high lepton momentum,
the $K_{\mu2}$ decay with a mis-identified muon ($E/p>0.95$, see
Fig.~\ref{fig:ke2-km2-kinematic-separation}c) is the largest
background source. The dominant process leading to
mis-identification of the muon as a positron is `catastrophic'
bremsstrahlung in or in front of the LKr leading to significant
energy deposit in the LKr. Mis-identification due to accidental LKr
clusters associated with the muon track is negligible, as concluded
from a study of the sidebands of track-cluster time difference and
distance distributions.

The muon mis-identification probability $P_{\mu e}$ has been
measured as a function of momentum. To collect a muon sample free
from the typical $\sim10^{-4}$ positron contamination due to $\mu\to
e$ decays, a $9.2X_0$ thick lead (Pb) wall covering $\sim 20\%$ of
the geometric acceptance was installed approximately 1.2~m in front
of the LKr calorimeter (between the two HOD planes) during a
dedicated period of data taking with $K^+$ and $K^-$ beams. The
$K_{e2}$ sample collected with the Pb wall installed is not used for
the $R_K$ measurement. The component from positrons which traverse
the Pb wall and are mis-identified as muons from $K_{\mu2}$ decay
with $p>30$~GeV/$c$ and $E/p>0.95$ is suppressed down to a
negligible level ($\sim 10^{-8}$) by energy losses in the Pb.

However, muon passage through the Pb wall affects the measured
$P_{\mu e}^\mathrm{Pb}$ via two principal effects: 1) ionization
energy loss in Pb decreases $P_{\mu e}$ and dominates at low
momentum; 2) bremsstrahlung in Pb increases $P_{\mu e}$ and
dominates at high momentum. To evaluate the correction factor
$f_\mathrm{Pb}=P_{\mu e}/P_{\mu e}^\mathrm{Pb}$, a dedicated MC
simulation based on Geant4 (version 9.2)~\cite{geant4} has been
developed to describe the propagation of muons downstream from the
last DCH, involving all electromagnetic processes including muon
bremsstrahlung~\cite{ke97}.

\begin{figure}[tb]
\begin{center}
\resizebox{0.50\textwidth}{!}{\includegraphics{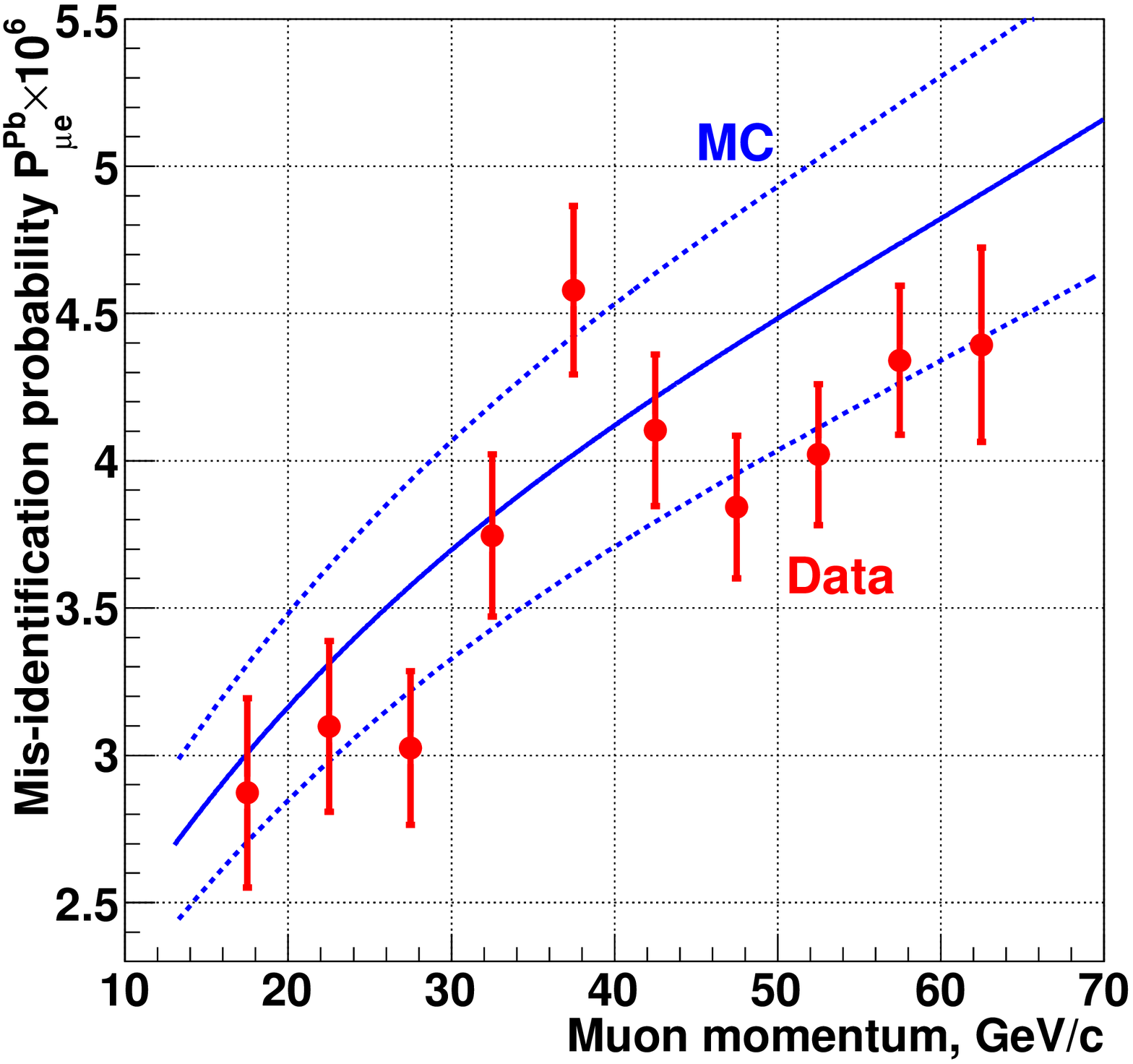}}%
\resizebox{0.50\textwidth}{!}{\includegraphics{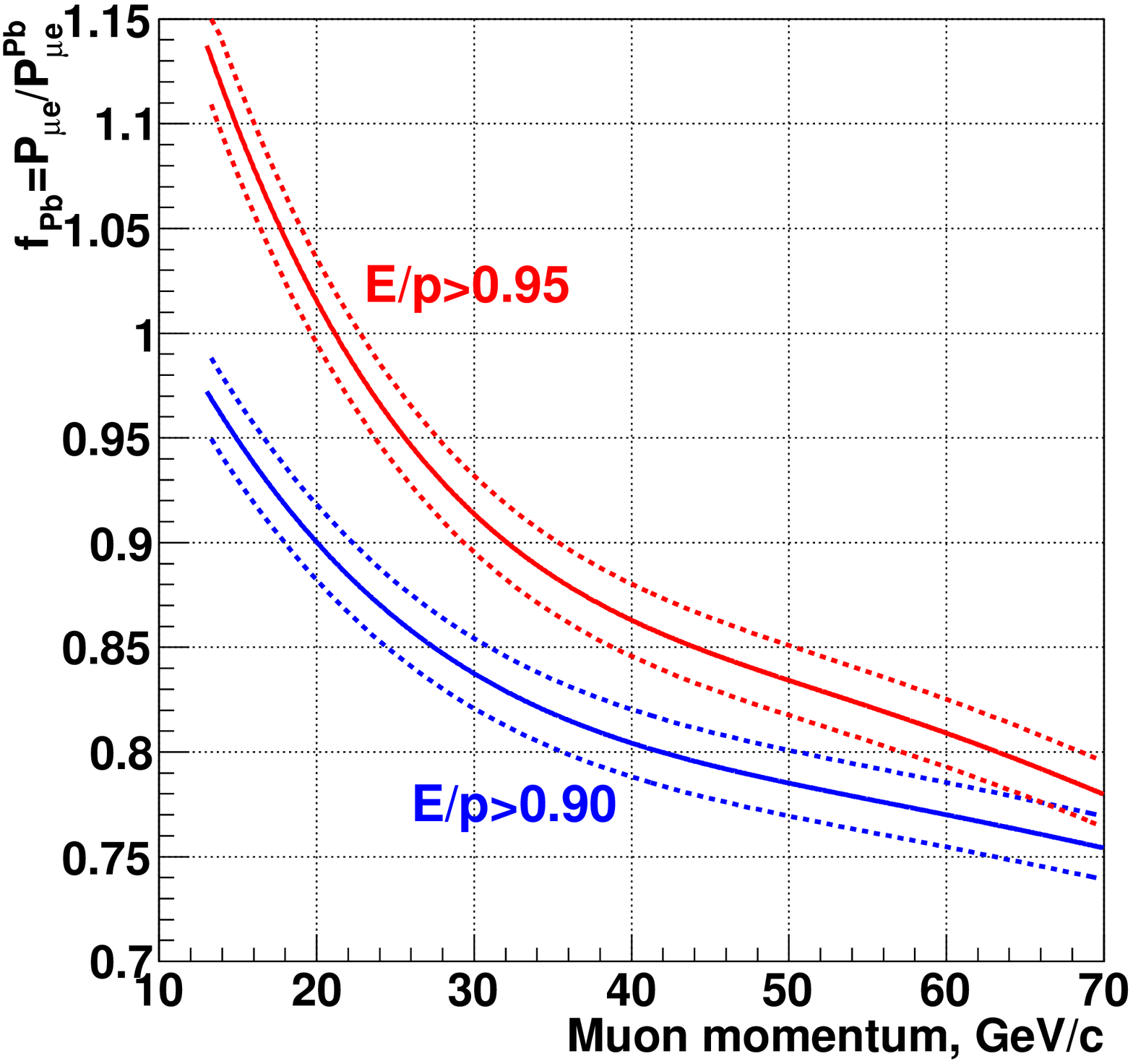}}
\put(-255,196){\bf\large (a)} \put(-30,196){\bf\large (b)}
\end{center}
\vspace{-12mm} \caption{(a) Mis-identification probability for muons
traversing the lead wall, $P_{\mu e}^\mathrm{Pb}$, for
$(E/p)_\mathrm{min}=0.95$ as a function of momentum: measurement
(solid circles with error bars) and simulation (solid line). (b)
Correction factors $f_\mathrm{Pb}=P_{\mu e}/P_{\mu e}^\mathrm{Pb}$
for the considered values of $(E/p)_\mathrm{min}$ , as evaluated
with simulation. Dotted lines in both plots indicate the estimated
systematic uncertainties of the simulation.} \label{fig:pmue}
\end{figure}

The measurements of $P_{\mu e}^{\mathrm{Pb}}$ in momentum bins
compared with the results of the MC simulation and the correction
factors $f_\mathrm{Pb}$ obtained from simulation, along with the
estimated systematic uncertainties of the simulated values, are
shown in Fig.~\ref{fig:pmue}.  The relative systematic uncertainties
on $P_{\mu e}$ and $P_{\mu e}^{\mathrm{Pb}}$ obtained by simulation
have been estimated to be $10\%$, and are mainly due to the
simulation of cluster reconstruction and energy calibration. However
the error of the ratio $f_\mathrm{Pb}=P_{\mu e}/P_{\mu
e}^\mathrm{Pb}$ is significantly smaller ($\delta
f_\mathrm{Pb}/f_\mathrm{Pb}=2\%$) due to cancellation of the main
systematic effects. The measured $P_{\mu e}^{\mathrm{Pb}}$ is in
agreement with the simulation within their uncertainties.

The positive correlation between the reconstructed
$M_\mathrm{miss}^2(e)$ and $E/p$, which are both computed using the
reconstructed track momentum, leads to an apparent dependence of
$P_{\mu e}$ on $M_\mathrm{miss}^2(e)$. This effect is significant
for intermediate lepton momenta where the $K_{\mu 2}$ background
comes from events with underestimated $M_\mathrm{miss}^2(e)$ and a
smaller muon mis-identification probability (see
Fig.~\ref{fig:ke2-km2-kinematic-separation}a). This correlation has
been taken into account.

The $K_{\mu 2}$ background contamination integrated over lepton
momentum has been computed to be $(6.11\pm0.22)\%$ using the
measured $P_{\mu e}^\mathrm{Pb}$ corrected by $f_\mathrm{Pb}$. The
quoted error comes from the limited size of the data sample used to
measure $P_{\mu e}^\mathrm{Pb}$ (0.16\%), the uncertainty $\delta
f_\mathrm{Pb}$ (0.12\%), and the model-dependence of the correction
for the $M_\mathrm{miss}^2(e)$ vs $E/p$ correlation (0.08\%). The
first error component is uncorrelated between the lepton momentum
bins, while the others are fully correlated.

As a stability check, the evaluation of $P_{\mu e}$ has been
performed with an additional requirement that the energy deposit in
the HOD counters downstream from the Pb wall is small (limited to
the equivalent of 1.5 to 3 minimum ionizing particles), which
strongly suppresses muons undergoing bremsstrahlung in the Pb wall.
The stability of $P_{\mu e}$ is consistent with the assigned
uncertainty $\delta f_\mathrm{Pb}$. Additionally, a stability check
of $R_K$ with respect to variation of $(E/p)_{\rm min}$ for lepton
momentum $p>25$~GeV/$c$ in the range from 0.90 to 0.97 has been
performed. The observed relative stability of $R_K$ within $\pm
0.2\%$, although the $K_{\mu 2}$ background varies from 17\% to 3\%,
is consistent with the uncertainty assigned to the $K_{\mu 2}$
background.

The $K_{\mu2}$ decay also contributes to background via $\mu^+\to
e^+\nu\bar\nu$ decays in flight. Energetic forward secondary
positrons compatible with $K_{e2}$ kinematics and topology are
suppressed by muon polarisation effects~\cite{mi50}. Radiative
corrections to the muon decay~\cite{ar02} lead to a further
$\sim10\%$ relative background suppression. This background
contamination has been estimated to be $(0.27\pm0.04)\%$, where the
dominant uncertainty is due to the simulated statistics.

\boldmath \subsubsection{$K^+\to e^+\nu\gamma$ background}
\unboldmath

$R_K$ is defined to be fully inclusive of internal bremsstrahlung
(IB) radiation~\cite{ci07}. The structure-dependent (SD) $K^+\to
e^+\nu\gamma$ process~\cite{bi93,ch08} may lead to a $K_{e2}$
signature if the positron is energetic and the photon is undetected.
In particular, the $\mathrm{SD}^+$ component with positive photon
helicity peaks at high positron momentum in the $K^+$ rest frame
($E^*_e\approx M_K/2$) and has a similar branching ratio to
$K_{e2}$. The background due to $K^+\to
e^+\nu\gamma~(\mathrm{SD}^-)$ decay with negative photon helicity
peaking at $E^*_e\approx M_K/4$ and the interference between the IB
and SD processes are negligible.

The $\mathrm{SD}^+$ background contribution has been estimated by MC
simulation as $(1.07\pm0.05)\%$, using a recent measurement of the
$K^+\to e^+\nu\gamma~(\mathrm{SD}^+)$ differential decay
rate~\cite{am09}. The quoted uncertainty is due to the limited
precision on the form factors and decay rate, and is therefore
correlated between lepton momentum bins. A stability check of $R_K$
against variation of the $E_\mathrm{veto}$ limit in a wide range has
been performed. While the $K^+\to e^+\nu\gamma~(\mathrm{SD^+})$
background is enhanced by a factor of 4.5 for
$E_\mathrm{veto}=14$~GeV with respect to $E_\mathrm{veto}=2$~GeV,
$R_K$ remains stable within $\pm 0.1\%$, which is consistent with
the above uncertainty.

\boldmath
\subsubsection{$K^+\to\pi^0 e^+\nu$ and $K^+\to\pi^+\pi^0$ backgrounds}
\unboldmath

The $K^+\to\pi^0e^+\nu$ decay produces a $K_{e2}$ signature if the
only reconstructed particle is an $e^+$ from $K^+$ or $\pi^0$ Dalitz
($\pi^0_D\to\gamma e^+e^-$) decays. The $K^+\to\pi^+\pi^0$ decay
leads to a $K_{e2}$ signature if the only reconstructed particle is
a $\pi^+$ mis-identified as $e^+$, or an $e^+$ from a
$\pi^0_D\to\gamma e^+e^-$ decay. The pion mis-identification
probability ($0.95<E/p<1.1$) has been measured to be
$(0.41\pm0.02)\%$ in the relevant momentum range from samples of
$K^+\to\pi^+\pi^0$ and $K^0_L\to\pi^\pm e^\mp\nu$ decays (the latter
collected during a special run).

Kinematically, $K^+\to\pi^0e^+\nu$ and $K^+\to\pi^+\pi^0$ decays can
be reconstructed with low missing mass in the $K_{e2}$ signal
region, either because the charged track undergoes a large multiple
scatter or because the kaon is in the high momentum tail of the beam
distribution. The systematic uncertainties due to subtraction of
these backgrounds have been estimated as 50\% of the contributions
themselves, due to the limited precision of the simulation of the
kaon momentum-distribution tails. The backgrounds are at a level
below 0.1\%.

\subsubsection{Beam halo background}
\label{sec:halo}

As no tracking is available in the beam region to tag an incoming
kaon, beam halo muons can become a source of background to $K_{e2}$
decays in case of $\mu^+\to e^+\nu_e\bar{\nu}_\mu$ decay or muon
mis-identification as a positron. The choice of the signal region in
terms of the longitudinal position of the kaon decay vertex has been
dictated by the kinematic distribution of this background (which
peaks in the upstream part of the vacuum volume).

The halo background has been measured directly by reconstructing the
$K^+_{e2}$ candidates from one control data sample collected with
the $K^-$ beam transmitted by the beam line and the $K^+$ beam (but
not its halo) blocked, and another control data sample collected
with both $K^+$ and $K^-$ beams blocked. The control samples are
normalised to the data in the region
$-0.3<M_\mathrm{miss}^2(\mu)<-0.1$ $(\mathrm{GeV}/c^2)^2$ populated
mainly by beam halo events. The `cross-talk' probability to
reconstruct a $K_{e2}^+$ candidate due to a $K^-$ decay with $e^+$
emission ($K^-\to\pi^0_D\ell^-\nu$, $K^-\to\pi^-\pi^0_D$,
$K^-\to\ell^-\nu e^+e^-$ with $\ell=e,\mu$) is at the level of $\sim
10^{-4}$ and is taken into account. The halo background rate and
kinematical distribution are qualitatively reproduced by a
simulation of the beam line.

The background contamination has been estimated to be
$(1.16\pm0.06)\%$, where the error comes from the limited size of
the control samples (uncorrelated between lepton momentum bins) and
the normalisation uncertainty due to decays of beam kaons and pions
upstream of the decay volume (correlated between momentum bins).

\boldmath
\subsubsection{Summary of backgrounds in the $K_{e2}$ sample}
\unboldmath

\begin{figure}[p]
\begin{center}
\resizebox{0.5\textwidth}{!}{\includegraphics{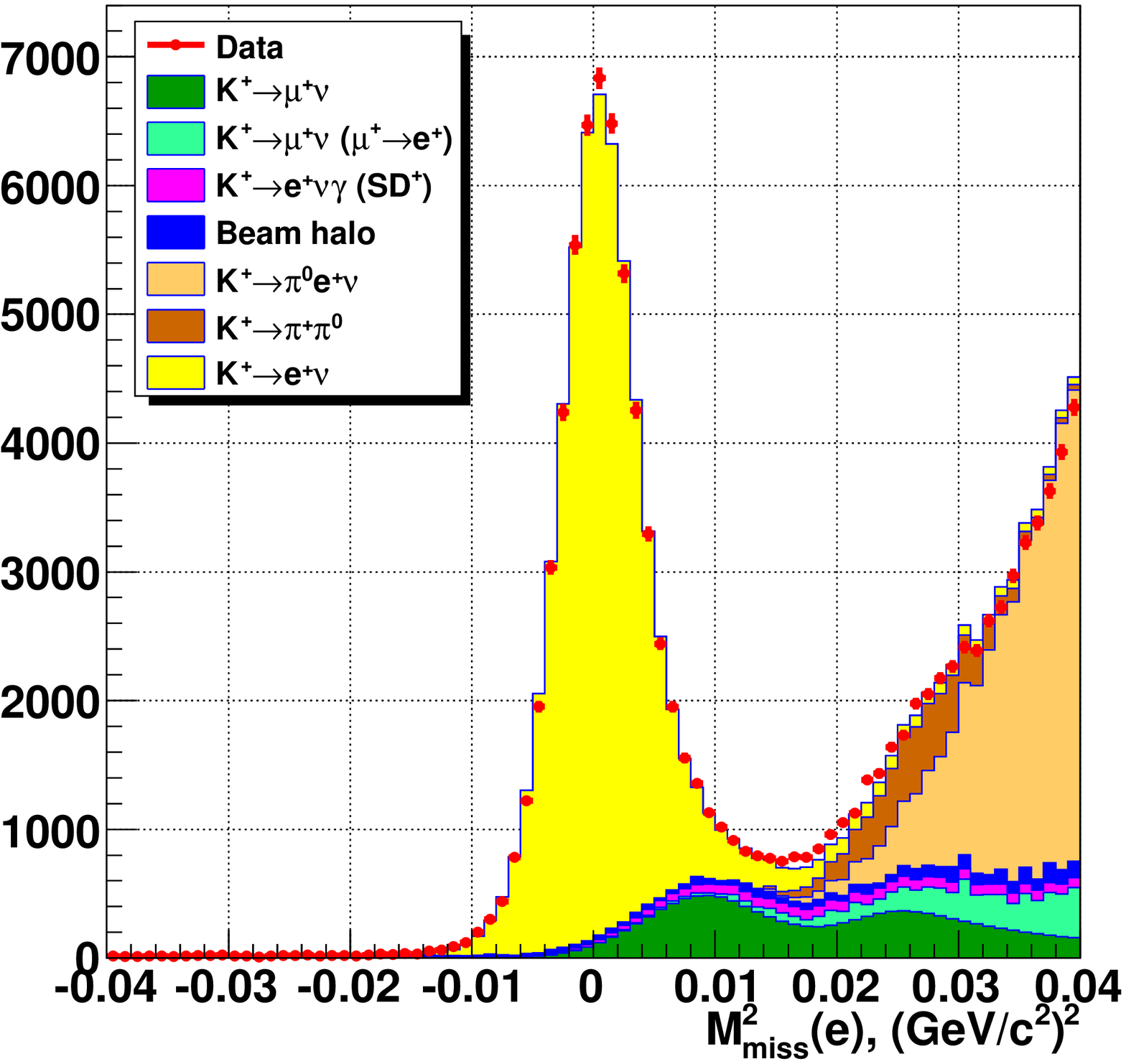}}%
\resizebox{0.5\textwidth}{!}{\includegraphics{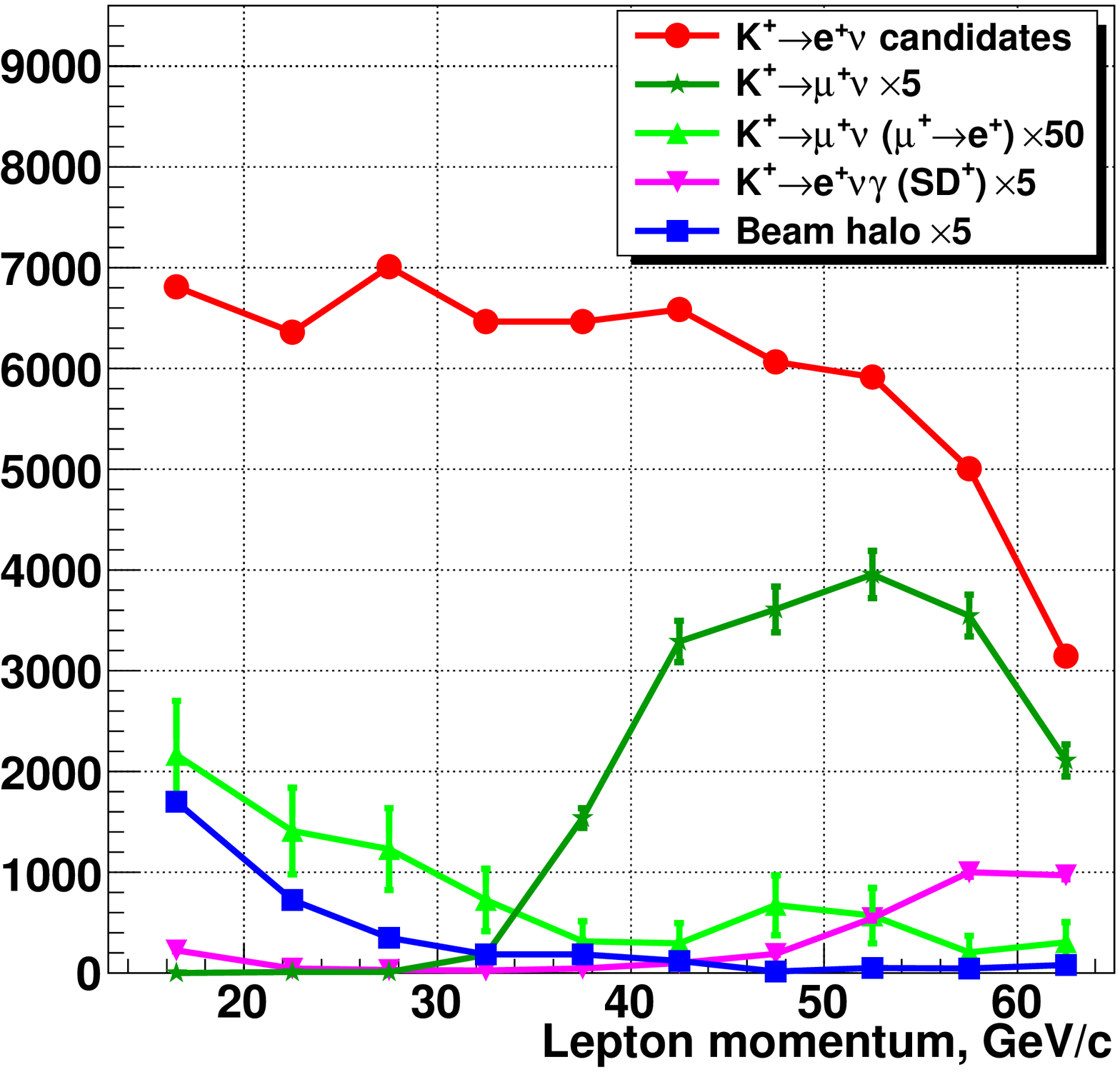}}
\put(-257,200){\bf\large (a)} \put(-200,200){\bf\large (b)}
\end{center}
\vspace{-12mm} \caption{(a) Reconstructed squared missing mass
$M_{\mathrm{miss}}^2(e)$ distribution of the $K_{e2}$ candidates
compared with the sum of normalised estimated signal and background
components. The small discrepancy between data and MC at low
$M_{\mathrm{miss}}^2(e)$ is due to the limited precision of MC beam
description, which is taken into account by systematic uncertainty
due to the acceptance correction. (b) Lepton momentum distributions
of the $K_{e2}$ candidates and the dominant backgrounds; the
backgrounds are scaled for visibility.} \label{fig:mm2e}
\end{figure}

Backgrounds integrated over lepton momentum are summarised in
Table~\ref{tab:bkg}. The total background contamination is
$(8.71\pm0.24)$\%; its uncertainty is smaller than the relative
statistical uncertainty of 0.43\%. The $M_{\mathrm{miss}}^2(e)$ and
lepton momentum distributions of $K_{e2}$ candidates and backgrounds
are shown in Fig.~\ref{fig:mm2e}.

\begin{table}[tb]
\begin{center}
\caption{Summary of backgrounds in the $K_{e2}$ sample.}
\label{tab:bkg} \vspace{2mm}
\begin{tabular}{lc}
\hline Source                  & $N_B/N(K_{e2})$\\
\hline $K_{\mu 2}$             & $(6.11\pm0.22)\%$\\
$K_{\mu 2}(\mu\to e)$          & $(0.27\pm0.04)\%$\\
$K^+\to e^+\nu\gamma~(\mathrm{SD}^+)$ & $(1.07\pm0.05)\%$\\
$K^+\to\pi^0 e^+\nu$           & $(0.05\pm0.03)\%$\\
$K^+\to\pi^+\pi^0$             & $(0.05\pm0.03)\%$\\
Beam halo                      & $(1.16\pm0.06)\%$\\
\hline
Total                          & $(8.71\pm0.24)\%$\\
\hline
\end{tabular}
\vspace{-4mm}
\end{center}
\end{table}

\boldmath
\subsection{The $K_{\mu2}$ sample}
\unboldmath

\begin{figure}[p]
\begin{center}
\resizebox{0.5\textwidth}{!}{\includegraphics{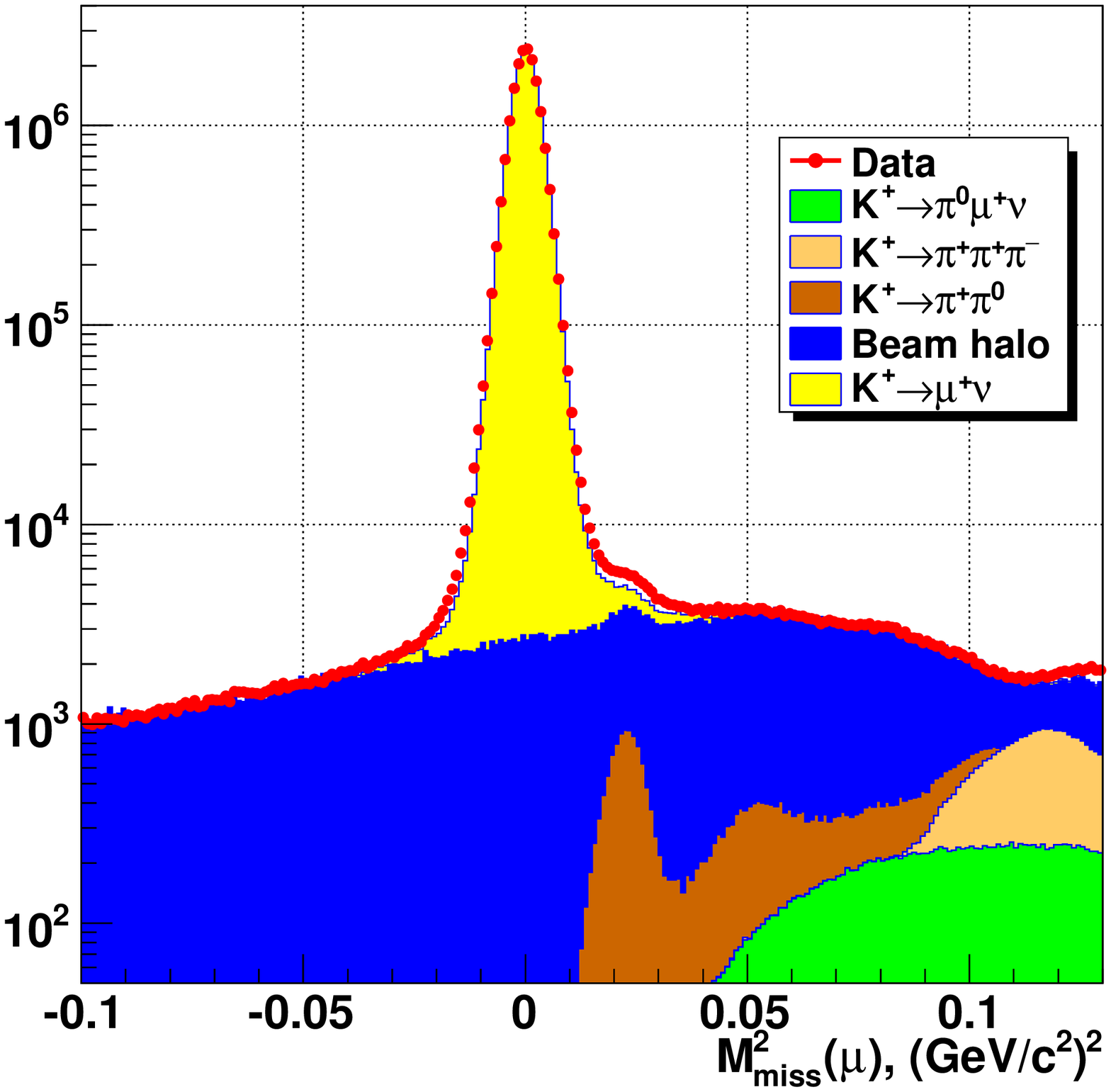}}%
\resizebox{0.5\textwidth}{!}{\includegraphics{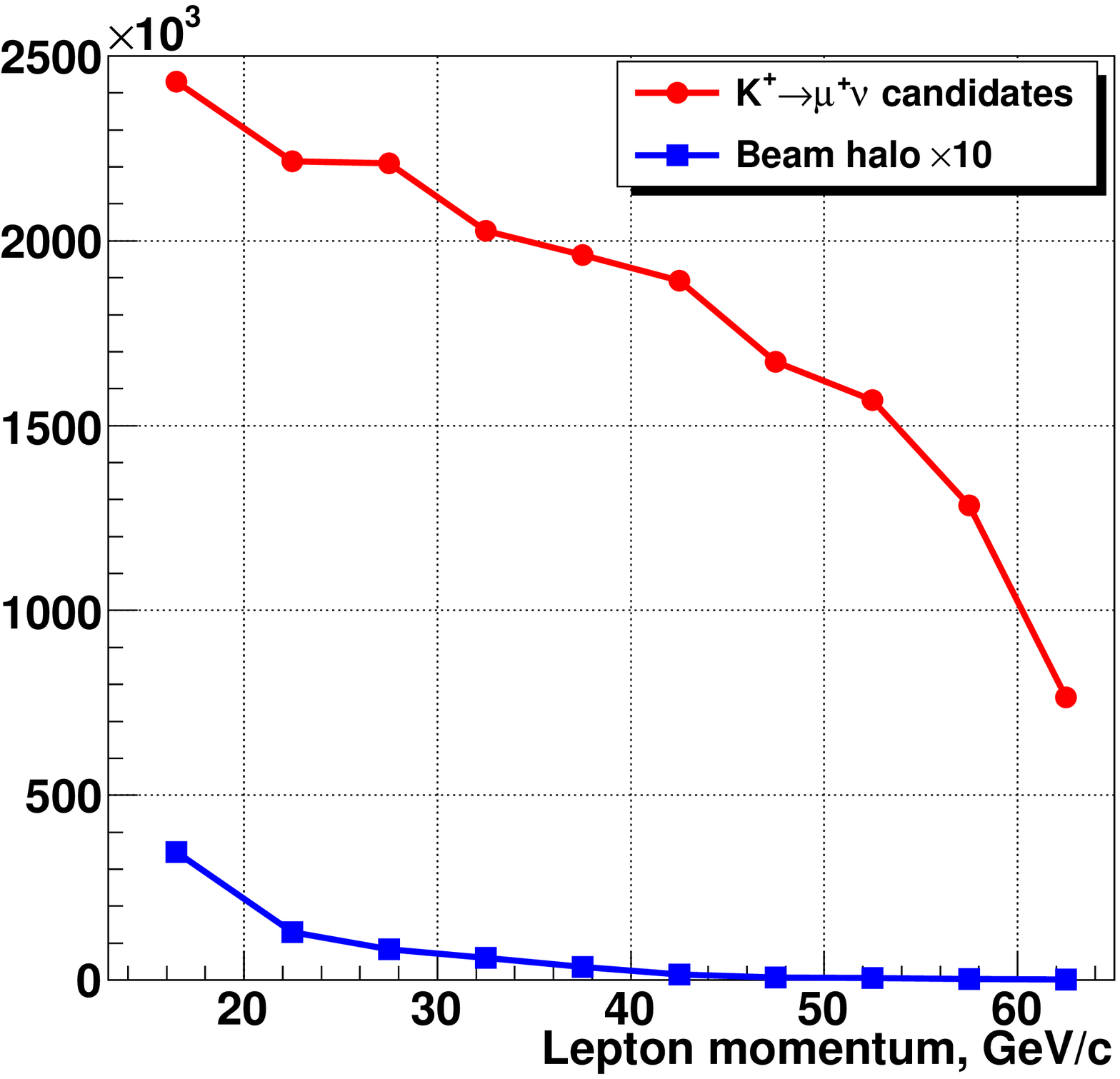}}
\put(-257,200){\bf\large (a)} \put(-200,180){\bf\large (b)}
\end{center}
\vspace{-12mm} \caption{(a) Reconstructed squared missing mass
$M_{\mathrm{miss}}^2(\mu)$ distribution of the $K_{\mu 2}$
candidates compared with the sum of normalised estimated signal and
background components. The deficit of reconstructed MC events in the
region of the $K^+\to\pi^+\pi^0$ peak is due to the limited
precision of the beam simulation, and is mostly outside the signal
region. (b) Lepton momentum distributions of the $K_{\mu2}$
candidates and the beam halo background (the latter is scaled for
visibility).} \label{fig:mm2m}
\end{figure}

The number of $K_{\mu 2}$ candidates collected with a trigger chain
involving downscaling by a factor of 150 is $N(K_{\mu
2})=1.803\times 10^7$. The only significant background source in the
$K_{\mu 2}$ sample is the beam halo. Its contribution is mainly at
low muon momentum, and has been measured to be $(0.38\pm0.01)\%$
using the same technique as for the $K_{e2}$ sample. The
$M_\mathrm{miss}^2(\mu)$ and muon momentum spectra of $K_{\mu2}$
candidates and the halo background are presented in
Fig.~\ref{fig:mm2m}.

\subsection{Geometrical acceptance correction}

The ratio of geometric acceptances $A(K_{\mu2})/A(K_{e2})$ in each
lepton momentum bin has been evaluated with MC simulation. The
radiative $K^+\to e^+\nu\gamma$ (IB) process, which is responsible
for the loss of about 5\% of the $K_{e2}$ acceptance by increasing
the reconstructed $M_{\rm miss}^2(e)$, is taken into account
following~\cite{bi93}, with higher order corrections according
to~\cite{we65,ga06}.

The acceptance correction is strongly influenced by bremsstrahlung
suffered by the positron in the material upstream of the
spectrometer magnet (Kevlar window, helium, DCHs). This results in
an almost momentum-independent loss of $K_{e2}$ acceptance of about
6\%, mainly by increasing the reconstructed $M_{\rm miss}^2(e)$. The
relevant material thickness has been measured by studying the
spectra and rates of bremsstrahlung photons produced by low
intensity 25~GeV/$c$ and 40~GeV/$c$ electron and positron beams
steered into the DCH acceptance, using special data samples
collected in the same setup by the NA48/2 experiment in 2004 and
2006. Using these measurements, the material thickness during the
2007 run has been estimated to be $(1.56\pm0.03)\% X_0$. The quoted
uncertainty is dominated by the limited knowledge of helium purity
in the spectrometer tank; its measured purity of $(92\pm4)\%$
corresponds to a thickness of $(0.26\pm0.03)\% X_0$. This translates
into a systematic uncertainty on $R_K$.

\begin{figure}[tb]
\begin{center}
\resizebox{0.5\textwidth}{!}{\includegraphics{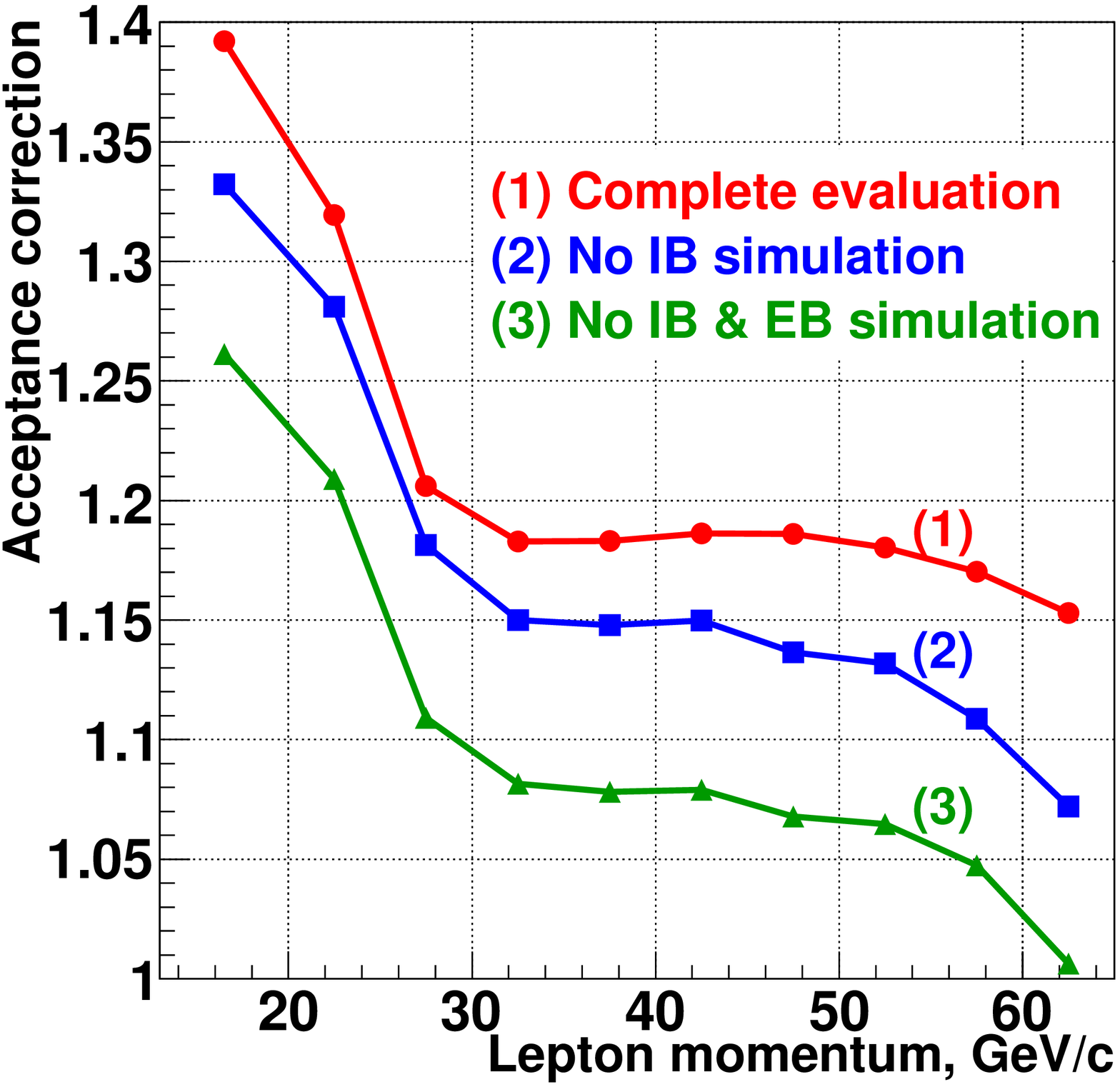}}%
\resizebox{0.5\textwidth}{!}{\includegraphics{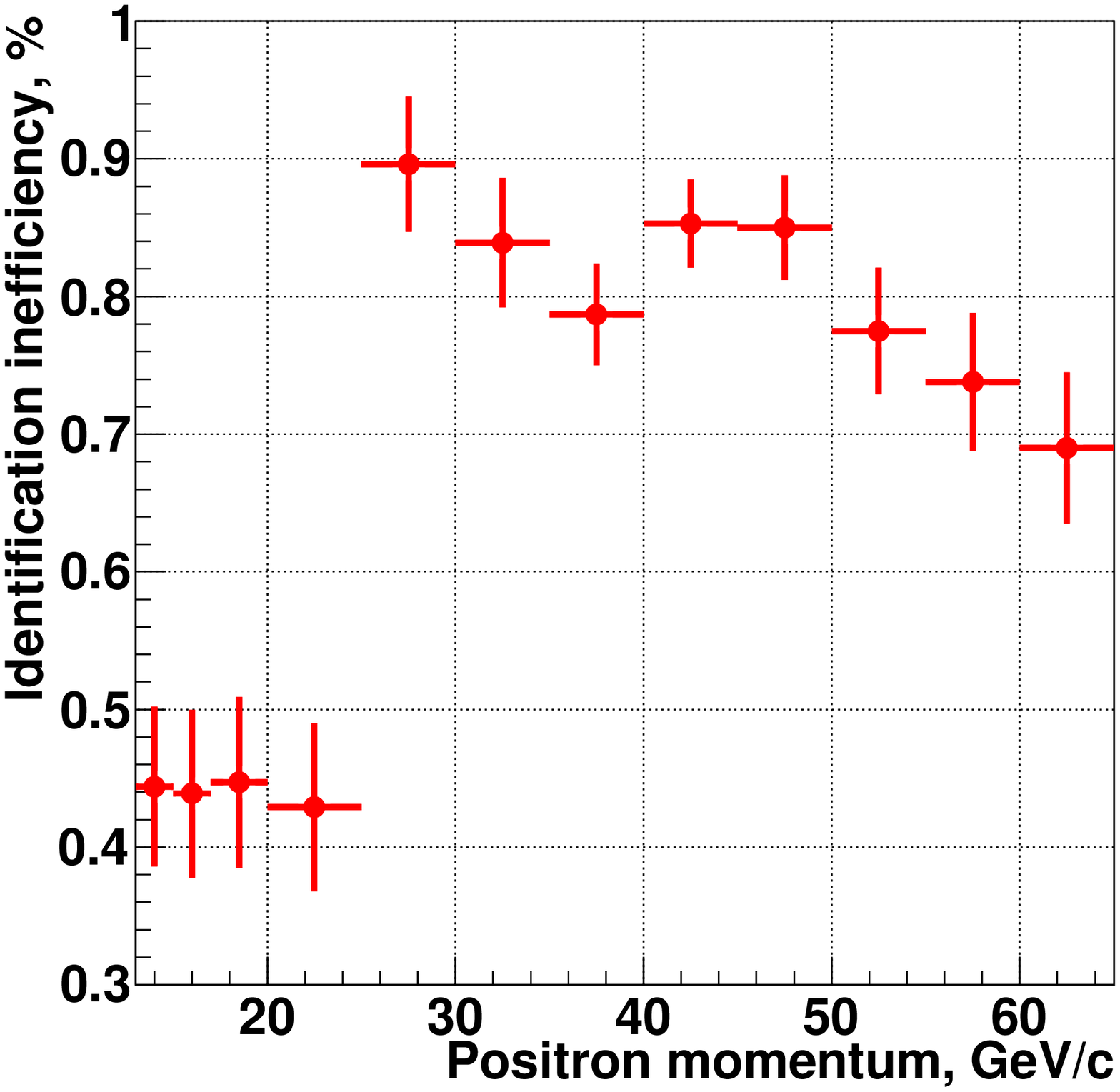}}
\put(-251,197){\bf\large (a)} \put(-25,197){\bf\large (b)}
\end{center}
\vspace{-12mm} \caption{(a) The acceptance correction
$A(K_{\mu2})/A(K_{e2})$ in lepton momentum bins; the corrections
neglecting internal (IB) and external (EB) bremsstrahlung radiation
are also presented. (b) The measured positron identification
inefficiency $1-f_e$ in lepton momentum bins; uncertainties in bins
are partially correlated. The lower inefficiency for $p<25$~GeV/$c$
is due to the relaxed positron identification requirement discussed
in Section~\ref{sec:selection}.} \label{fig:acc-id}
\end{figure}

The acceptance correction $A(K_{\mu2})/A(K_{e2})$ in lepton momentum
bins is presented in Fig.~\ref{fig:acc-id}a. The corrections
evaluated without internal (IB) and external (EB) bremsstrahlung
radiation are also presented to illustrate the magnitudes of the
corresponding effects. The correction is enhanced at low lepton
momentum because the radial distributions of positrons from $K_{e2}$
decays in the DCH planes are wider than those of muons from $K_{\mu
2}$ decays, and low momentum leptons are not fully contained within
the geometric acceptance due to the limited transverse sizes of the
DCHs.

The track reconstruction inefficiency due to interactions in
spectrometer material is included into the acceptance correction.
Simulation of the positron track reconstruction inefficiency (which
is $\sim10^{-3}$ in the analysis track momentum range) has been
validated with a sample of $K^+\to\pi^+\pi^0_D$ decays. The muon
track reconstruction inefficiency evaluated with MC simulation is
$\sim2\times 10^{-4}$. Systematic effects due to imperfect
simulation of the reconstruction efficiency are negligible.

Apart from helium purity, the main sources of systematic uncertainty
of the acceptance correction are the limited knowledge of beam
profile and divergence, and the simulation of soft radiative
photons. A separate uncertainty has been assigned to account for the
finite precision of the DCH alignment.

\subsection{Lepton identification efficiencies}

The $E/p$ ratio provides powerful particle identification criteria.
The momentum-dependent positron identification window $(E/p)_{\rm
min} < E/p < 1.1$
includes more than 99$\%$ of the $K_{e2}$ events, while suppressing
muons by a factor of $1/P_{\mu e}\sim10^6$. The requirement
$E/p<0.85$ leads to a negligible inefficiency of the muon
identification.

A pure sample of $4\times 10^7$ positrons, selected kinematically
from $K^+ \to \pi^0 e^+ \nu$ (charged $K_{e3}$) decays collected
with the $K_{e2}$ trigger concurrently with the main $K_{\ell 2}$
data set, is used to calibrate the energy response of each LKr cell
and to study $f_e$ with respect to local position and time
stability. However, the momentum range of the positrons from charged
$K_{e3}$ decays is kinematically limited, preventing a sufficiently
precise measurement of $f_e$ above $50~\mathrm{GeV}/c$. Therefore a
dedicated data sample was recorded in a special 15 hour long run
with a broad momentum band $K^0_L$ beam. Electrons and positrons
from the $4\times10^{6}$ collected $K^0_L \to \pi^\pm e^\mp \nu$
(neutral $K_{e3}$) decays allow the determination of $f_e$ in the
whole analysis momentum range.

The measurements of $f_e$ have been performed in bins of lepton
momentum; a finer binning is used inside the lowest of the 10
standard bins to improve the determination of local inefficiencies,
which peak at low momentum. Separate measurements have been
performed for several identified groups of LKr cells with higher
local inefficiencies. Efficiency measurements with the charged and
neutral $K_{e3}$ decays agree to better than 0.1\%.
Fig.~\ref{fig:acc-id}b shows the measurements of $1-f_e$ in momentum
bins used to evaluate corrections to $R_K$, obtained as the weighted
mean of charged and neutral $K_{e3}$ measurements for momenta up to
$50~\mathrm{GeV}/c$, and as neutral kaon measurements for higher
momenta. The inefficiency averaged over the $K_{e2}$ sample is
$1-f_e = (0.73\pm0.05)\%$, where the uncertainty takes into account
the statistical precision and the small differences between charged
and neutral kaon results.

\subsection{Trigger and readout efficiencies}

The efficiency of the $Q_1$ trigger condition has been measured
using $K_{\mu2}$ events triggered with a control LKr signal. The
inefficiency integrated over the $K_{\mu2}$ sample is
$(1.4\pm0.1)\%$. As a consequence of its geometric uniformity and
the similarity of the $K_{e2}$ and $K_{\mu2}$ distributions over the
HOD plane, it nearly cancels between the $K_{e2}$ and $K_{\mu2}$
samples, and the residual systematic bias is negligible. The
inefficiency of the 1-track condition also largely cancels in the
ratio $R_K$, but is anyway negligible.

Thus the trigger efficiency correction
$\epsilon(K_{\mu2})/\epsilon(K_{e2})$ is determined by the
efficiency $\epsilon(E_\mathrm{LKr})$ of the LKr energy deposit
trigger signal $E_\mathrm{LKr}>10$~GeV entering the $K_{e2}$ trigger
chain only. The inefficiency $1-\epsilon(E_\mathrm{LKr})$ is only
significant in the lowest lepton momentum bin of (13, 20)~GeV/$c$,
which is close to the trigger energy threshold and is thus affected
by the online energy resolution. A sample of events triggered with a
control $Q_1$ signal passing all $K_{e2}$ selection criteria except
the $M_{\rm miss}^2(e)$ constraint, therefore dominated by $K_{e3}$
events with two lost photons, has been used to measure
$1-\epsilon(E_\mathrm{LKr})$ in the lowest momentum bin to be
$(0.41\pm0.05_{\rm stat.})\%$. Corrected for the difference of
positron distributions in the LKr plane between the $K_{e2}$ sample
and the control sample, it translates into
$1-\epsilon(E_\mathrm{LKr}) = (0.61\pm0.20)\%$ for the $K_{e2}$
sample. The correction and its uncertainty are significant due to
the presence of several locally inefficient regions. The resulting
uncertainty on $R_K$ is negligible.

Energetic photons not reconstructed in the LKr may initiate showers
by interacting in the DCHs or the beam pipe, causing the DCH hit
multiplicities to exceed the limits allowed by the 1-track trigger
condition. Among the backgrounds, only the $K^+\to
e^+\nu\gamma~(\mathrm{SD}^+)$ receives a non-negligible correction
due to the 1-track inefficiency. The inefficiency for
$K^+\to\pi^0e^+\nu$ events with two lost photons has been measured
to vary in the range from 0.1 to 0.3 depending on track momentum.
The extrapolation of this result to $K^+\to
e^+\nu\gamma~(\mathrm{SD}^+)$ with one lost photon relies on
simulation. The corresponding uncertainty has been propagated into
$R_K$.

The global LKr readout inefficiency, affecting the $K_{e2}$
reconstruction only, has been measured using an independent readout
system to be $1-f_\mathrm{LKr}=(0.20\pm0.03)\%$, stable in time.
This measurement has been cross-checked, with limited precision, by
a study of the LKr response in a sample of $\pi^0_{DD}\to4e^\pm$
decays reconstructed from spectrometer information only.

\section{Result and discussion}

\begin{figure}[t]
\begin{center}
\resizebox{0.5\textwidth}{!}{\includegraphics{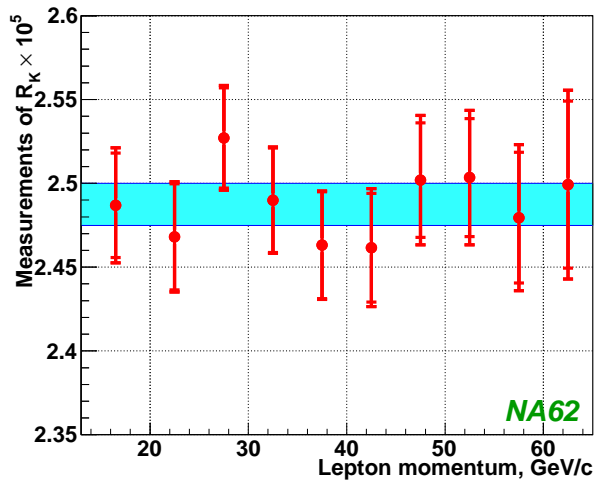}}%
\end{center}
\vspace{-12mm} \caption{Measurements of $R_K$ in lepton momentum
bins with their uncorrelated statistical uncertainties and the
partially correlated total uncertainties. The average $R_K$ and its
total uncertainty are indicated by a band.} \label{fig:rk}
\end{figure}

A $\chi^2$ fit to the measurements of $R_K$ in the 10 lepton
momentum bins has been performed, taking into account the bin-to-bin
correlations between the systematic errors. To validate the assigned
systematic uncertainties, extensive stability checks have been
performed in bins of kinematic variables and by varying selection
criteria and analysis procedures. The fit result is
\begin{equation}
R_K = (2.487\pm 0.011_{\mathrm{stat.}}\pm
0.007_{\mathrm{syst.}})\times 10^{-5} =(2.487\pm0.013)\times
10^{-5},
\end{equation}
with $\chi^2/{\rm ndf}=3.6/9$. The individual measurements with
their statistical and total uncertainties, and the combined result
are displayed in Fig.~\ref{fig:rk}. The uncertainties of the
combined result are summarised in Table~\ref{tab:err}.

This is the most precise $R_K$ measurement to date. It is consistent
with the KLOE measurement~\cite{am09} and the SM expectation
$R_K^\mathrm{SM}=(2.477\pm0.001)\times 10^{-5}$, and can be used to
constrain  multi-Higgs~\cite{ma06} and fourth generation~\cite{la10}
new physics scenarios. The experimental accuracy is still an order
of magnitude behind the SM accuracy, which motivates further
precision measurements of $R_K$.


\begin{table}
\begin{center}
\caption{Summary of the uncertainties on $R_K$.}
\label{tab:err}\vspace{2mm}
\begin{tabular}{lc}
\hline
Source & $\delta R_K\times 10^5$\\
\hline
Statistical                         & 0.011  \\
\hline
~~~$K_{\mu2}$ background               & 0.005  \\
~~~$K^+\to e^+\nu\gamma~(\textrm{SD}^+)$ background & 0.001\\
~~~$K^+\to\pi^0 e^+\nu$, $K^+\to\pi^+\pi^0$ backgrounds & 0.001\\
~~~Beam halo background                & 0.001\\
~~~Helium purity                       & 0.003\\
~~~Acceptance correction               & 0.002\\
~~~Spectrometer alignment              & 0.001\\
~~~Positron identification efficiency  & 0.001\\
~~~1-track trigger efficiency          & 0.002\\
~~~LKr readout inefficiency            & 0.001\\
Total systematic                       & 0.007\\
\hline
Total & 0.013\\
\hline
\end{tabular}
\vspace{-1cm}
\end{center}
\end{table}


\section*{Acknowledgements}

It is a pleasure to express our appreciation to the staff of the
CERN laboratory and the technical staff of the participating
laboratories and universities for their efforts in operation of the
experiment and data processing. We gratefully acknowledge the
excellent performance of the BlueBEAR computing facility at the
University of Birmingham, where the simulations have been performed.
The IHEP and INR groups have been supported in part by the RFBR
grants N08-02-91016 and N10-02-00330. We are grateful to Vincenzo
Cirigliano, Gino Isidori and Paride Paradisi for valuable
discussions.



\end{document}

%% file: na62-authors-2007.tex
\begin{center}
{\Large The NA62 collaboration}\\
\vspace{2mm}
 C.~Lazzeroni,
 A.~Romano\\
{\em \small University of Birmingham, Edgbaston, Birmingham,
B15 2TT, United Kingdom} \\[0.2cm]
 A.~Ceccucci,
 H.~Danielsson,
 V.~Falaleev,
 L.~Gatignon,
 S.~Goy Lopez$\,$\footnotemark[1],
 B.~Hallgren$\,$\footnotemark[2],
 A.~Maier,
 A.~Peters,
 M.~Piccini$\,$\footnotemark[3],
 P.~Riedler\\
{\em \small CERN, CH-1211 Gen\`eve 23, Switzerland}
% ...
\\[0.2cm]
 T.~Numao,
 F.~Retiere \\
{\em \small TRIUMF, 4004 Wesbrook Mall, Vancouver, British Columbia, V6T 2A3, Canada} \\[0.2cm]
\end{center}
%
%
%
\footnotetext[1]{Present address: CIEMAT, E-28040 Madrid, Spain}
\footnotetext[2]{Present address: University of Birmingham,
Edgbaston, Birmingham, B15 2TT, United Kingdom}
\footnotetext[3]{Present address: Sezione dell'INFN di Perugia,
I-06100 Perugia, Italy}
\footnotetext[4]{Present address: CERN, CH-1211 Gen\`eve 23,
Switzerland}
\footnotetext[5]{Also at Dipartimento di Fisica, Universit\`a di
Modena e Reggio Emilia, I-41100 Modena, Italy}
\footnotetext[6]{Also at Dipartimento di Fisica, Universit\`a di
Firenze, I-50019 Sesto Fiorentino (FI), Italy}
\footnotetext[7]{Also at Istituto di Fisica,
Universit\`a di Urbino, I-61029 Urbino, Italy}
\footnotetext[8]{Present address: CP3, Universit\'e catholique de
Louvain, B-1348 Louvain-la-Neuve, Belgium}
\footnotetext[9]{Also at University of Birmingham, Edgbaston,
Birmingham, B15 2TT, United Kingdom}
\footnotetext[10]{Present address: Dipartimento di Fisica
Sperimentale dell'Universit\`a di Torino, I-10125 Torino, Italy}
\footnotetext[11]{Present address: University College Dublin School
of Physics, Belfield, Dublin 4, Ireland}

%% file: ke2_.bbl
\begin{thebibliography}{99}
%
\bibitem{ci07}
V. Cirigliano and I. Rosell, Phys. Rev. Lett. {\bf 99} (2007)
231801.
%
%
%
%
\bibitem{ma06}
A. Masiero, P. Paradisi and R. Petronzio, Phys. Rev. {\bf D74}
(2006) 011701.
%
\bibitem{ma08}
A. Masiero, P. Paradisi and R. Petronzio, JHEP {\bf 0811} (2008)
042.
%
\bibitem{el09}
J. Ellis, S. Lola and M. Raidal, Nucl. Phys. {\bf B812} (2009) 128.
%
\bibitem{la10}
H. Lacker and A. Menzel, JHEP {\bf 1007} (2010) 006.
%
\bibitem{cl72}
A.G. Clark {\it et al.}, Phys. Rev. Lett. {\bf 29} (1972) 1274.
%
\bibitem{he75}
K.S. Heard {\it et al.}, Phys. Lett. {\bf B55} (1975) 327.
%
\bibitem{he76}
J. Heintze {\it et al.}, Phys. Lett. {\bf B60} (1976) 302.
%
\bibitem{pdg}
K. Nakamura {\it et al.} (PDG), J. Phys. {\bf G37} (2010) 075021.
%
\bibitem{am09}
F. Ambrosino {\it et al.}, Eur. Phys. J. {\bf C64} (2009) 627;
ibid. {\bf C65} (2010) 703.
%
\bibitem{fa07}
V. Fanti {\it et al.}, Nucl. Instrum. Methods {\bf A574}
(2007) 433.
%
\bibitem{ba07}
R. Batley {\it et al.}, Eur. Phys. J. {\bf C52} (2007) 875.
%
\bibitem{be95}
D. B\'ed\'er\`ede {\it et al.}, Nucl. Instrum. Methods {\bf A367}
(1995) 88.
%
\bibitem{ba96}
G. Barr {\it et al.}, Nucl. Instrum. Methods {\bf A370} (1996) 413.
%
%
\bibitem{geant3} GEANT Description and Simulation Tool, CERN
Program Library Long Writeup {\bf W5013} (1994).
%
\bibitem{geant4}
S. Agostinelli {\it et al.}, Nucl. Instrum. Methods {\bf A506} (2003) 250.
%
\bibitem{ke97}
S.R. Kelner, R.P. Kokoulin and A.A. Petrukhin, Phys. Atom. Nucl. {\bf 60}
(1997) 576.
%
\bibitem{mi50}
L. Michel, Proc. Phys. Soc. {\bf A63} (1950) 514.
%
\bibitem{ar02}
A. Arbuzov, A. Czarnecki and A. Gaponenko, Phys. Rev. {\bf D65}
(2002) 113006.
%
\bibitem{bi93}
J. Bijnens, G. Ecker and J. Gasser, Nucl. Phys. {\bf B396} (1993)
81.
%
\bibitem{ch08}
C.H. Chen, C.Q. Geng and C.C. Lih, Phys. Rev. {\bf D77} (2008)
014004.
%
\bibitem{we65}
S. Weinberg, Phys. Rev. {\bf 140} (1965) B516.
%
\bibitem{ga06}
C. Gatti, Eur. Phys. J. {\bf C45} (2006) 417.
%
\end{thebibliography}
